# The Importance of UV Capabilities for Identifying Inhabited Exoplanets with Next Generation Space Telescopes

Edward Schwieterman*[1,2,3,4], Christopher Reinhard[3,5], Stephanie Olson[1,3], Timothy Lyons[1,3]

[1]*Department of Earth Sciences, University of California, Riverside, California*
[2]*NASA Postdoctoral Program, Universities Space Research Association, Columbia, Maryland*
[3]*NASA Astrobiology Institute, Alternative Earths Team, Riverside, California*
[4]*Blue Marble Space Institute of Science, Seattle, Washington*
[5]*School of Earth and Atmospheric Sciences, Georgia Institute of Technology, Atlanta, Georgia*

**ABSTRACT**

The strongest remotely detectable signature of life on our planet today is the photosynthetically produced oxygen ($O_2$) in our atmosphere. However, recent studies of Earth's geochemical proxy record suggest that for all but the last ~500 million years, atmospheric $O_2$ would have been undetectable to a remote observer—and thus a potential false negative for life. During an extended period in Earth's middle history (2.0 – 0.7 billion years ago, Ga), $O_2$ was likely present but in low concentrations, with $pO_2$ estimates of ~ 0.1 – 1% of present day levels. Although $O_2$ has a weak spectral impact at these low abundances, $O_3$ in photochemical equilibrium with that $O_2$ would produce notable spectral features in the UV Hartley-Huggins band (~0.25 µm), with a weaker impact in the mid-IR band near 9.7 µm. Thus, taking Earth history as an informative example, there likely exists a category of exoplanets for which conventional biosignatures can only be identified in the UV. In this paper, we emphasize the importance of UV capabilities in the design of future space-based direct imaging telescopes such as HabEx or LUVOIR to detect $O_3$ on planets with intermediate oxygenation states. We also discuss strategies for mitigating against 'false positives'—that is, $O_3$ produced by abiotic processes. More generally, this specific example highlights the broad implications of studying Earth history as a window into understanding potential exoplanet biosignatures.

## 1. Introduction and Relevance

The search for life beyond our solar system is a prominent goal within the NASA astrobiology program, emphasized in both the 2008 Astrobiology Roadmap and the 2015 Astrobiology Strategy. The rapid evolution of exoplanet science from detection to characterization studies and the discovery of planets in the habitable zones of nearby stars underscores the timeliness of this effort. The nearest and best chance for identifying life on exoplanets will be provided by large (30-m class) ground-based observatories and future 10-m class space-based direct-imaging telescopes. While the James Webb Space Telescope (JWST), set to launch in 2019, will provide an unprecedented opportunity to characterize exoplanets through phase curves, secondary eclipse observations, and transit transmission spectroscopy, space-based direct-imaging characterization of terrestrial exoplanets will have to wait for dedicated observatories such as the LUVOIR or HabEx concepts (e.g., Mennesson et al. 2016; Bolcar et al. 2017). The science and technology definition teams (STDTs) for both concepts are convening now, and as was the case for JWST, **broad determination of the required instrument capabilities for these missions will be made many years and perhaps decades before their launch dates**. It is therefore essential to accurately and swiftly identify the minimum capabilities for a direct-imaging observatory to accomplish top level objectives such as identifying inhabited planets (e.g., Stark et al., 2014).

The most commonly referenced biosignature gases are $O_2$ and its photochemical byproduct $O_3$, due to $O_2$'s exclusive biological production on Earth through oxygenic photosynthesis and the strong thermodynamic and kinetic disequilibrium it produces in the atmosphere (Des Marais et al., 2002). Alternative biosignature gases, surface signatures, and overarching frameworks have been proposed and should remain an important part of the conversation (see reviews in Schwieterman et al., 2018; Meadows et al., 2018, Catling et al., 2018, Walker et al., 2018, and Fujii et al., 2018); however it remains important to fully benchmark and examine $O_2/O_3$ signatures. We assert that Earth's history tells us that $O_3$, best detected in the UV, is a more sensitive and consistent indicator of planetary scale photosynthetic life than $O_2$, thus minimizing the potential for false negatives. In section 2 below we review Earth's oxygenation history as context for this assertion. In section 3 we examine the remote detectability of $O_2$ and $O_3$ through that history, and we discuss mitigation against false positives in section 4. We summarize our recommendations in section 5.

## 2. Earth's Oxygenation History

Although molecular oxygen ($O_2$) currently represents ~20% of Earth's atmospheric mass, the amount of $O_2$ in our atmosphere has evolved dramatically over time. Indeed, for the vast majority of Earth history, atmospheric $O_2$ levels were orders of magnitude below those characteristic of the modern Earth. During Archean time (3.8 – 2.5 billion years ago, Ga), the preservation of non-mass-dependent sulfur isotope anomalies in marine sediments fingerprints atmospheric $O_2$ concentrations well below $10^{-5}$ times the present atmospheric level (PAL; Farquhar et al., 2001; Pavlov and Kasting, 2002; Claire et al., 2006). The disappearance of sulfur isotope anomalies from Earth's rock record at ~2.3 Ga points to a rise in atmospheric $O_2$ (Luo et al., 2016), but a number of geochemical archives suggest extended periods of very low atmospheric $O_2$ well after this initial rise (Lyons et al., 2014; Planavsky et al., 2014; Cole et al., 2016). It is thus possible that atmospheric $O_2$ has been well below ~1% of the modern value for as much as 90% of Earth's evolutionary history. Just as the Archean Earth has been presented as an analog for Earth-like exoplanets (Arney et al., 2016), the subsequent Proterozoic eon (2.5 – 0.5 Ga), comparable in duration, provides an additional template for understanding the potential atmospheric states of habitable exoplanets—and the fundamental controls that should determine the evolving redox states for many complex planetary systems.

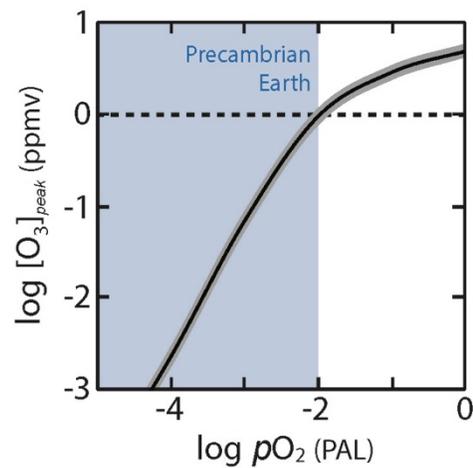

**Figure 1 – $O_3$ abundance as a function of $pO_2$.** Calculation of peak stratospheric $O_3$ as a function of ground-level $O_2$ based results from Kasting and Donahue (1980). From Reinhard et al. (2017).

## 3. Remote detectability of $O_2/O_3$ throughout Earth's history

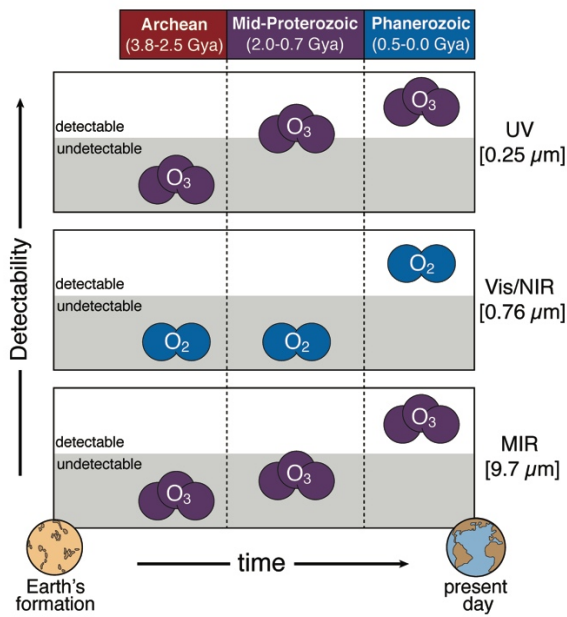

**Figure 2 – Conceptual illustration of $O_2/O_3$ detectability in the UV, Vis/NIR, and MIR through Earth history.** *This is a simplified representation of the spectral data in Figure 3. Adapted from Schwieterman et al. (2018).*

Molecular oxygen ($O_2$) shows no significant spectral features at mid-IR wavelengths but absorbs strongly at the Fraunhofer A and B bands (0.76 and 0.69 μm, respectively) and at 1.27 μm. The most prominent of these features is the Fraunhofer A band, but this feature is expected to have appreciable depth only at atmospheric levels of ~1% PAL or higher (Des Marais et al., 2002; Segura et al., 2003). As a result, direct detection and/or quantification of $O_2$ would have been extremely challenging for all but the last ~500 million years of Earth's history (Reinhard et al., 2017).

However, $O_2$ can be detected by proxy through searching for signs of atmospheric $O_3$. On Earth, $O_3$ is produced in the stratosphere through photolysis of $O_2$ and recombination of O atoms with ambient $O_2$ molecules through the Chapman reactions. In addition, photochemical models demonstrate that the atmospheric abundance of $O_3$ shows strong dependence on atmospheric $O_2$ at oxygenation states significantly below modern values (see Figure 1), with the result that atmospheric $O_3$ levels are potentially a very sensitive indicator of surface $O_2$ production on terrestrial planets with low to intermediate oxygen levels compared to those present today.

Ozone has a number of significant spectral features at UV, visible, and IR wavelengths. In particular, $O_3$ absorbs strongly within the Hartley-Huggins bands at ~0.35-0.2 μm and the Chappuis bands between 0.5 and 0.7 μm and shows an additional strong absorption feature at 9.7 μm. **From the standpoint of detection, it is the near-UV Hartley-Huggins feature centered at ~0.25 μm that is most important, because it is sensitive to extremely low levels of atmospheric $O_3$**. This feature saturates at peak $O_3$ abundances of less than ~1 ppmv, corresponding to a background atmospheric $O_2$ level of around 1% PAL (Reinhard et al., 2017). This critical observation indicates that it may have been possible to fingerprint the presence of biogenic $O_2$ in the atmosphere using the Hartley-Huggins feature of $O_3$ for, more than half of Earth's evolution, despite background $O_2$ levels that would have rendered direct detection of molecular $O_2$ extremely difficult. Figure 2 shows a schematic, simplified representation of the relationship between $O_2$ and $O_3$ concentrations and conservative estimated detection thresholds during three eons of Earth history (Archean, Proterozoic, and Phanerozoic/modern). Figure 3 displays simulated spectral observations of the UV Hartley-Huggins band, the $O_2$-A band, and the 9.7 μm $O_3$ band for upper and lower estimates of the concentrations of these gases during each eon (data obtained from Table 1 of Reinhard et al., 2017). From Figure 3 it is apparent that the **most sensitive indicator of atmospheric $O_2$ is the UV $O_3$ band**, which would have created a measurable impact on Earth's spectrum for perhaps ~50% of its history, versus ~10% for $O_2$.

### 4. Mitigating against 'false positives'

Recent work has illustrated several scenarios for abiotic buildup of $O_2$ and $O_3$ in planetary atmospheres, such as through extensive hydrogen escape or robust $CO_2$ photolysis (see reviews in Meadows, 2017 and Meadows et al., 2018). One relevant observation is that in most cases, potentially detectable abiotic $O_3$ is more easily generated than detectable $O_2$ (e.g., Domagal-Goldman et al., 2014). However, the most compelling scenarios for 'false positive' $O_2/O_3$ biosignatures concern planets orbiting M-dwarf stars, which possess extended pre-main sequence phases (enhancing the probability of hydrogen loss and $O_2$ buildup) and high FUV/NUV flux ratios (enhancing the photolysis rate of O-bearing molecules such as $CO_2$). Fortunately, however, inner working angle (IWA) constraints for direct-imaging telescopes will favor the angular separation of habitable zone planets orbiting early K, G, and F type stars, where the processes that may produce abiotic $O_2/O_3$ are disfavored. In addition, the absence of certain UV/Vis/NIR spectral indicators, such as $O_4$ and CO, can help rule out these 'false positive' mechanisms (e.g., Schwieterman et al., 2016). The most plausible mechanisms for abiotic $O_2/O_3$ in planets orbiting solar-type stars is steady H-escape from thin, water-rich atmospheres lacking in non-condensing gases (Wordsworth & Pierrehumbert, 2014), in which case blue-near-UV wavelength capabilities will be important for estimating atmospheric mass through characterizing Rayleigh scattering. This positive dynamic is enhanced further by the higher photospheric temperatures of FGK stars, generating more near-UV flux and thus greater S/N at wavelengths relevant to $O_3$ characterization. Of course, assessing the host star's UV spectrum would also help directly constrain plausible photolysis rates and the resulting potential for abiotic $O_2/O_3$ (e.g., France et al., 2016). Thus, 'false positives' can be successfully mitigated by both target selection and multi-wavelength characterization of planet and star, **which would be aided by UV capability**. Further, mitigating against $O_2$ 'false negatives' requires UV, or less effectively, MIR wavelengths inaccessible to the HabEx/LUVOIR concepts.

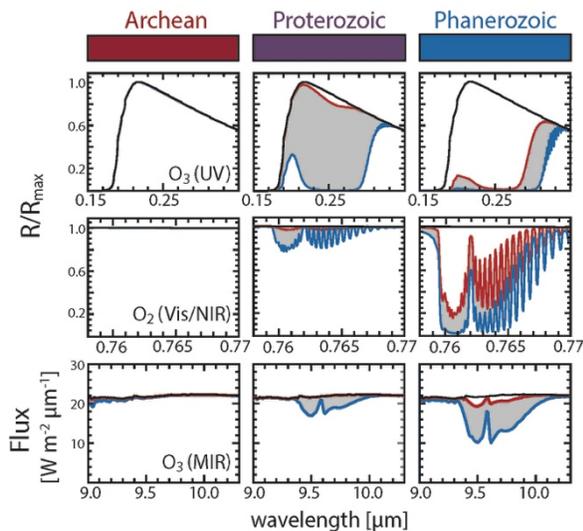

**Figure 3 – Spectral features of $O_2$ and $O_3$ in the UV, Vis/NIR, and MIR.** *Modeled planet spectra, at 1 $cm^{-1}$ resolution, of the $O_3$ Huggins-Hartley band (0.25 μm), $O_2$-A band (0.76 μm), and 9.7 μm $O_3$ band at different geologic times assuming gas abundances informed by biogeochemical modelling and geochemical proxy constraints. Generated with the SMART radiative transfer model (Meadows & Crisp, 1996).*

## 5. Discussion and Conclusions

Remote observations of Earth would have failed to detect $O_2$ for 90% of its history if limited to optical and near-infrared wavelengths. In contrast, sensitivity to UV wavelengths would have allowed the detection of $O_3$, thus fingerprinting the presence of $O_2$ in our atmosphere for half its lifetime. There is no guarantee that habitable or inhabited exoplanets will be like Earth or recapitulate its atmospheric evolution, but if we take our planet as an informative example, **it is clear that detection thresholds of $pO_2 > 1\%$ PAL or higher could eliminate the potential for life detection on planets with intermediate oxygenation states ($10^{-5}$ PAL $< pO_2 < 1\%$ PAL). Future work should carefully combine simulated planetary spectra and realistic instrumental performance for space telescopes with UV capabilities**. As a supporting proof of concept, the LCROSS mission has detected $O_3$ in remote observations of Earth (Robinson et al., 2014). Additionally, UV wavelengths provide more favorable IWA requirements than optical or near-infrared observations for both coronagraph- and starshade-based designs (Seager et al., 2015; Robinson et al., 2016), allowing a greater number of planets to be surveyed and a likely larger biosignature yield (e.g., Stark et al., 2014). Importantly, 'false positive' $O_3$ biosignatures can be mitigated through target selection and multi-wavelength planetary characterization (including the UV), while $O_2$ 'false negatives' cannot be eliminated without the UV.